\documentclass{ws-ijmpd}
\usepackage[super,compress]{cite}
\begin{document}

\markboth{J. Grain}
{The perturbed universe in the deformed algebra approach of LQC}

%
\catchline{}{}{}{}{}
%

\title{THE PERTURBED UNIVERSE IN THE DEFORMED ALGEBRA APPROACH OF LOOP QUANTUM COSMOLOGY}

\author{JULIEN GRAIN}

\address{CNRS, Institut d'Astrophysique Spatiale, UMR8617\\
Orsay, F-91405, France\\
and \\
Universit\'e Paris Sud \\
Orsay, F-91405, France\\
julien.grain@ias.u-psud.fr}

\maketitle


\begin{abstract}
Loop quantum cosmology is a tentative approach to model the universe down to the Planck era where quantum gravity settings are needed. The quantization of the universe as a dynamical space-time is inspired by Loop Quantum Gravity ideas. In addition, loop quantum cosmology could bridge contact with astronomical observations, and thus potentially investigate quantum cosmology modellings in the light of observations. To do so however, modelling both the background evolution and its perturbations is needed. The latter describe cosmic inhomogeneities that are the main cosmological observables. In this context, we present the so-called {\it deformed algebra} approach implementing the quantum corrections to the perturbed universe at an effective level by taking great care of gauge issues. We particularly highlight that in this framework, the algebra of hypersurface deformation receives quantum corrections, and we discuss their meaning. The primordial power spectra of scalar and tensor inhomogeneities are then presented, assuming initial conditions are set in the contracting phase preceding the quantum bounce and the well-known expanding phase of the cosmic history. These spectra are subsequently propagated to angular power spectra of the anisotropies of the cosmic microwave background. It is then shown that regardless of the choice for the initial conditions inside the effective approach for the background evolution (except that they are set in the contracting phase), the predicted angular power spectra of the polarized $B$-modes exceed the upper bound currently set by observations. The exclusion of this specific version of loop quantum cosmology establishes the falsifiability of the approach, though one shall not conclude here that either loop quantum cosmology or loop quantum gravity is excluded.
\end{abstract}


\ccode{PACS numbers: 98.80.Qc, 98.80.Jk, 04.60.Bc, 04.60.Ds, 04.60.Pp,}

\section{Introduction}
\label{sec:intro}
The building of a consistent quantum theory of space-times is one of the most outstanding problems in contemporary physics. This problem is in fact two-fold. Not only its mathematical construction is made difficult, but quantum gravity also appears to be out of reach of any possible experiments because the Planck scale at which quantum gravitational effects should pop up is about thirteen orders of magnitude beyond the energy scales probed at the LHC. The situation for an observational probe of quantum gravity is however not hopeless, as shown by many recent results in Loop Quantum Cosmology (LQC) \cite{barrau_2014}, and we detailed here an example of this.

The strongest argument in favour of quantum gravity is the existence of singularities in the classical theory of general relativity \cite{hawking_1973}. In other words, general relativity predicts its own breakdown and a way out is mandatory. Quantum gravity then appears as a rather natural solution since the Planck scale is attained by approaching the singularities (at least for part of them). But this tells us more. Searching for experimental or observational signatures of quantum gravity can be phrased in the search of gravitational phenomena which are singular in the sense of general relativity. This has important consequences in cosmology: the standard $\Lambda$CDM model is based on the Friedmann-Lema\^\i{t}re-Robertson-Walker (FLRW) metric with matter and radiation contents that make the universe a singular space-time in its past. As a consequence, a complete description of our universe, including the Planck era, should incorporate quantum gravity settings. In the framework of LQC, it has been shown that the initial Big Bang singularity is replaced by a regular quantum bounce between a classical contracting phase and a classical expanding phase (introductory reviews can be found in e.g. Refs. \refcite{bojowald_2008,ashtekar_2003}, and detailed discussions of the background evolution can be found in Refs. \refcite{linsefors_2013,bolliet_2015}). The existence of the bounce is a direct consequence of the loop quantization: the use of {\it holonomies} of the connection as dynamical variables and the fact that the quantum spectrum of the area operator has a minimal gap provide a regularization of the curvature in the Planck era. The detailed dynamics of the classical contracting and expanding branches depends on the details of the matter content. If this content is a single massive scalar field, the quantum universe then enters in a phase of classical inflation after the bounce with a given probability, thus linking the quantum bounce to the more standard dynamics of the universe \cite{linsefors_2013,sloan_2010}. \\

Observing that our universe went through a {\it bounce} and not originated from a {\it bang} requires however additional inputs. LQC is a quantization of the symmetry-reduced FLRW space-time inspired by Loop Quantum Gravity (LQG) ideas. This leads to a quantum description of a strictly homogeneous and isotropic space-time. Any observer is however confined within the universe and one should rely on internal tracers of its dynamics, which are the cosmological inhomogeneities as the large-scale structures (galaxies, clusters of galaxies, filaments, etc.) or the anisotropies of the Cosmic Microwave Background (CMB). Cosmological inhomogeneities are  degrees of freedom of the underlying gravitational field coupled to the matter fields. They are modelled through perturbations of the FLRW space-time, that is by linearizing the classical Einstein's equations about the FLRW solution. Otherwise stated, our universe is not modelled by a strictly FLRW space-time, but by a {\it perturbed} FLRW space-time as a sufficient implementation of the {\it statistical} homogeneity and isotropy\footnote{We stress that a perturbed FLRW space-time is a sufficient conditions to model a statistically homogeneous and isotropic universe. This is however not a necessary one.}. In the context of LQC though, the gravitational field is not described anymore by the Einstein's field equations (at least in the deep quantum era close to the bounce), and the theory of cosmological perturbations should be amended first to take into account that they live in, and even originate from, a quantum space-time.

With the theory of quantum cosmological perturbations at hand, the procedure to compare predictions from an LQC modelling of the cosmic history to astronomical observations follows the approach commonly used in classical cosmology to probe inflation. In the latter context, cosmological inhomogeneities resulted from the gravitational amplification of the quantum fluctuations of the vacuum during the inflationary phase. For singe field inflation, there are two types of inhomogeneities: scalar modes, $\mathcal{R}$, which are perturbations of the 3-dimensional scalar curvature, and tensor modes, $h^i_a$, which are primordial gravitational waves. The standard predictions are the statistical properties of these inhomogeneities at the end of inflation\footnote{To be precise here, what is needed is the statistical properties of the inhomogeneities at the onset of the radiation-dominated era. These statistical properties have thus to be propagated through the phase of reheating which comes in between the end of inflation, and the radiation-dominated era in the primordial universe. There are different arguments showing that the statistical properties of the inhomogeneities are not distorted through reheating (at least in the linear perturbation theory) a part from the standard homogeneous and isotropic redshifting of scales. As a consequence, knowing the statistical properties of the inhomogeneities at the end of inflation is sufficient to know them at the begining of the radiation-dominated era.}. In the most simple scenario of single field inflation there is only very tiny deviations from gaussianities (as currently favored by observations\cite{planck_gauss}), as a consequence of the gaussian nature of the vacuum state and a quasi-linear evolution. Both kinds of inhomogeneities are thus mainly described by their 2-points correlation functions, or more commonly, by their primordial power spectra: $\mathcal{P}_\mathrm{S}(k)=\frac{k^3}{2\pi}\left<\mathcal{R}_k\mathcal{R}_k^\star\right>$, and $\mathcal{P}_\mathrm{T}(k)=(16Gk^3)\sum_{s=1,2}\left<h^i_a(k,s)h^a_i(k,s)\right>$, with a sum over the two helicity states for the tensor modes (note that the power spectrum only depends on the norm of the wavevector as a result of statistical isotropy). At the end of inflation, the universe is filled with inhomogeneities. The scalar modes then serve as primordial seeds for the formation of large-scale structures through gravitational collapse latter on in the cosmic history. They also leave their footprints in the form of anisotropies in the total intensity (or temperature) of the CMB, and in the $E$-modes of its linear polarization. The tensor modes also leave their imprints on the CMB. They are subdominant in intensity and $E$-modes but leave an (undetected yet) unique signature in the $B$-modes of the linear polarization at large angular scales \cite{zaldarriaga_1997}. The amplitude of the tensor modes is currently constrained by observations to be lower than about one tenth of the amplitude of the scalar modes at scales $k\sim0.002$ Mpc$^{-1}$\cite{planck_inf,planckbicep}. 

The key-point is that any constraints on the $\mathcal{P}_\mathrm{S(T)}$'s from e.g. the CMB, can be translated into constraints on inflation, that is on a period in the cosmic history about $10^{-30}$ second after the Big Bang. Similarly in LQC, cosmological perturbations evolves through the contraction and the bounce before inflation. Some distortions of the $\mathcal{P}_\mathrm{S(T)}$'s as compared to the standard prediction of inflation are thus legitimately expected. In other words, a clear inspection of the statistical properties of the primordial inhomogeneities through the observations of the CMB anisotropies, allows in principle for setting constraints on the cosmic history before inflation, meaning a quantum bounce in the specific case of LQC. \\

There is a consensus for the quantum background in LQC. However, the case of cosmological perturbations is more lively debated and different paths have been developed. These paths to a theory of quantum cosmological perturbations and the conceptual issues they raised are discussed in details in another article of this special issue \cite{barrau_ijmpd}. Here we will focus on the so-called deformed algebra approach \cite{barrau_jcap} with holonomy corrections (note that the deformed algebra has been extended to the case of inverse volume corrections \cite{iv_bojo,iv_tom}). The procedure to make contact between LQC predictions and astronomical observations remains however the same, meaning that in principle astronomical observations could be used to discriminate between different paths. 

This paper is organised as follows. Sec. \ref{sec:da} is dedicated to a detailed presentation of the deformed algebra approach. The primordial power spectra in this approach, and assuming that initial conditions for the perturbations are set in the classical contracting phase, are presented in Sec. \ref{sec:pk}. We then discuss their implications in terms of the angular power spectra of the CMB $B$-modes in Sec. \ref{sec:cl}. Finally, we conclude in Sec. \ref{sec:concl}. 

\section{An overview of the deformed algebra approach}
\label{sec:da}
\subsection{A brief description of the Hamiltonian formalism}
LQC is based on the Hamiltonian formulation of general relativity introduced by Arnowitt, Deser and Misner\cite{adm}, and then extended using the so-called Ashtekar variables\cite{ashtekar_1986}. A detailed presentation of the ADM formalism can be found in Ref. \refcite{thiemann}. This relies on the foliation of 4-dimensional (Riemannian) space-times into a set of 3-dimensional hypersurfaces, $\left(\Sigma_t\right)_{t\in\mathbb{R}}$. In this setting, the dynamics is given by the tim evolution of canonical variables defined on the hypersurfaces, that the evolution from one hypersurface to another. On hypersurfaces, the geometrical canonical variables are the induced metric $q_{ab}$ and its associated canonical momentum $P^{ab}=\delta S/\delta\dot{q}_{ab}$. This phase-space is extended to include the lapse function, $N$, and the shift vector, $N^a$, defining the foliation, and their associated momenta. The Einstein-Hilbert action does not depend on the time derivative of $N$ and $N^a$. This is so because chosing a foliation is arbitrary as a result of the diffeomorphism invariance of general relativity. The total Hamiltonian derived from the Einstein-Hilbert action is $\mathcal{H}\propto\int d^3x\left[\lambda C+\lambda^aC_a+\left|N\right| H+N^aH_a\right]$, with $C=\delta S/\delta\dot{N}$ and $C_a=\delta S/\delta \dot{N}^a$ which are constrained to be zero; $H$ and $H_a$ are the Hamiltonian and spatial diffeomorphism constraints that are functionals of $q_{ab}$ and $P^{ab}$. Because $C=0$ and $C_a=0$ should be preserved accross evolution, i.e. $\dot{C}=\left\{\mathcal{H},C\right\}=0$ and $\dot{C}_a=\left\{\mathcal{H},C_a\right\}=0$, this leads to $H=H_a=0$. General relativity is thus a totally constrained system, $\mathcal{H}\approx0$ on the solution of general relativity. Let us now look at the constraints surface, that is the subspace of the phase space for which the constraints hold. Introducing the smeared Hamiltonian and spatial diffeomorphism constraints, $S(N)=\int d^3xNH$ and $D(N^a)=\int d^3x N^aH_a$, there time evolution is given by $\left\{\mathcal{H},H\right\}$ and $\left\{\mathcal{H},D\right\}$ which can be shown to be equivalent to \cite{thiemann}:
\begin{eqnarray}
	\left\{D(M^a),D(N^a)\right\}&=&D\left(M^b\partial_bN^a-N^b\partial_bM^a\right), \\
	\left\{D(M^a),S(N)\right\}&=&S\left(M^b\partial_bN-N\partial_bM^b\right), \\
	\left\{S(M),S(N)\right\}&=&sD\left(q^{ab}(M\partial_bN-N\partial_bM)\right),	
\end{eqnarray}
with $s=1$ for the Lorentzian signature, and $s=-1$ for the Euclidean one. (Note that in the following we will systematically denotes $S$ and $D$ the smeared constraints associated to $H$ and $H_a$ respectively.) The above algebraic structure is called the {\it hypersurface deformation algebra} and it leads to $\left\{\mathcal{H}(N,N^a),\mathcal{H}(M,M^a)\right\}\approx0$. In other words, the constraint surface is preserved under evolution generated by the constraints. Since the lapse and the shift are mere Lagrange multipliers, the total Hamiltonian can thus be simplified to $\mathcal{H}\propto\int d^3x\left[\left|N\right| H+N^aH_a\right]$ and the dynamics it generates is equivalent to the Einstein's equations. The constraints $S$ and $D$ are said to form a system of {\it first-class constraints}.

We note that this framework has a geometrical interpretation. Once hypersurfaces are embedded in the Riemannian 4-dimensional space, one can defined how they can be deformed either by displacing a point within the hypersurface, or by displacing it along a geodesic of the 4-dimensional space which is initially orthogonal to the considered hypersurface \cite{hojman_1976}. These transformations are defined by the fact that the embedding space is a 4-dimensional Riemannian space. The generators of those two transformations, $\mathcal{D}_a$ and $\mathcal{D}$, satisfies the algebra of hypersurface deformation. This means that the Hamiltonian constraint and the spatial diffeomorphism constraint are representations of the transformations of the embedded hypersurfaces. On the space of solutions, the Hamiltonian constraint generates diffeomorphism orthogonal to $\Sigma_t$, and the spatial diffeomorphism constraint generates diffeormorphism which preserves $\Sigma_t$. \\

This formalism can be extended to the case of the Ashtekar variables. In this formalism, one introduces the densitized triads, $E^a_i$, defined by $\det(q_{ab})q^{ab}=E^a_iE^b_j\tilde{\delta}^{ij}$ ($\tilde\delta$ is the 3-dimensional flat metric). The densitized triad is derived from a triad valued in $su(2)$. There is obviously a redundancy here since any local rotation of the triads leads to the same induced metric, $q_{ab}$. Since rotations are generated by the Gauss constraint, this constraint will be additionally imposed in the Hamiltonian. The canonical variables associated to $E^a_i$ is the Ashtekar connection: $A^i_a=K^i_a+\gamma \Gamma^i_a$, with $K^i_a$ derived from the extrinsic curvature $K_{ab}$ through $K^i_a=K_{ab}E^{bi}/\sqrt{\det(q_{ab})}$, $\Gamma^i_a$ is derived from the spin-connection defining the transport of a 3-dimensional Lorentz vector on the hypersurfaces, and $\gamma$ is the Barbero-Immirzi parameter usually taken to be real-valued. The associated Poisson bracket is $\left\{A^i_a(\vec{x}),E^b_j(\vec{y})\right\}=8\pi G\gamma\delta^i_j\delta^b_a\delta^3(\vec{x}-\vec{y})$. Using this, general relativity is recovered from the following total Hamiltonian
\begin{equation}
	\mathcal{H}=\displaystyle\int_{\Sigma_t}d^3x\left(NC+N^aC_a+N^iC_i\right),
\end{equation}
with the Hamiltonian, spatial diffeomorphism, and Gauss constraints given by
\begin{eqnarray}
	C &=& \frac{1}{16\pi G} \frac{E^a_iE^b_j}{\sqrt{|\det E|}} \left[ {\epsilon^{ij}}_k F_{ab}^k -2(1+\gamma^2)K^i_{[a} K^j_{b]} \right], \\
	C_a &=& \frac{1}{8\pi G \gamma}(E^b_i F^i_{ab} - A^i_a C_i) \label{wiezDiff1}, \\
	C_i &=&\frac{1}{8\pi G \gamma}\mathcal{D}_a E^a_i = \frac{1}{8\pi G \gamma} \left(\partial_aE^a_i+\epsilon_{ijk} A^j_a E^a_k \right).
\end{eqnarray}
$F^i_{ab} =\partial_a A^i_b -\partial_b A^i_a+{\epsilon^i}_{jk} A^j_a A^k_b$ is the curvature of the Ashtekar connection. The situation is then rather similar to what was described for the ADM formalism. The total Hamiltonian is zero on the constraints surface. On defining the smeared constraints
\begin{eqnarray}
\mathcal{C}_1 &=& G[N^i] = \int_{\Sigma_t}d^3x\ N^i C_i, \\
\mathcal{C}_2 &=& D[N^a] =  \int_{\Sigma_t}d^3x\ N^a C_a, \\
\mathcal{C}_3 &=& S[N] =  \int_{\Sigma_t}d^3x\ N C,
\end{eqnarray}
one can check that they defined an algebra of first-class constraints
\begin{equation}
\{ \mathcal{C}_I, \mathcal{C}_J \} = {f^K}_{IJ}(A^j_b,E^a_i) \mathcal{C}_K. \label{algebra}
\end{equation}
This ensures that the constraints surface is preserved under time evolution. \\

\subsection{Cosmology in the Ashtekar formalism}
Considering first the case of a strictly homogeneous and isotropic space-time (with flat spatial metric for simplicity), the line element is given by the FLRW metric $ds^2=-dt^2+a^2(t)d\vec{x}\cdot d\vec{x}$. Homogeneity and isotropy greatly simplifies the Hamiltonian analysis. In the Ashtekar formalism, $A^i_a=\gamma \bar{k}(t)\delta^i_a$ and $E^a_i=\bar{p}(t)\delta^a_i$ (with $\bar{p}=a^2$), with the simplified Poisson bracket $\left\{\bar{k},\bar{p}\right\}=8\pi G/3$. Introducing a scalar field with a potential $V(\varphi)$, the Hamiltonian constraint reads $H=-3\sqrt{\bar{p}}\bar{k}^2/(8\pi G)+[\pi_\varphi^2/(2\bar{p}^{3/2})+\bar{p}^{3/2}V(\varphi)]$ ($\pi_\varphi$ is the momentum of the scalar field with the Poisson bracket $\left\{\varphi,\pi_\varphi\right\}=1$)\footnote{Note that $A^i_a=\bar{c}(t)\delta^i_a$ is more commonly used. The Poisson bracket is then $\left\{\bar{c},\bar{p}\right\}=8\pi\gamma G/3$, and the gravitational Hamiltonian constraint $H=-3\sqrt{\bar{p}}\bar{c}^2/(8\pi\gamma^2 G)$. Here we use $\bar{c}\equiv\gamma\bar{k}$ more common for implementing effective LQC corrections. The two are equivalent.}. Because of homogeneity and isotropy, the spatial diffeomorphism constraint is zero, and the total Hamiltonian is given by the Hamiltonian constraint. Using $\dot{f}=\left\{\mathcal{H},f\right\}$ (with $f$ being either the gravitational variables or the scalar field variables) as well as the constraint $\mathcal{H}=0$ allows for recovering the classical Friedmann and Raychaudhuri equations, and the Klein-Gordon equation for the scalar field. \\

An Hamiltonian treatment of the cosmological perturbations has been first proposed in Ref. \refcite{langlois_1994} using the ADM formalism, and a wrap-up of this using Ashtekar variables can be found in Refs. \refcite{cailleteau_2012a,cailleteau_2012}. Our presentation follows those two last references. We will not go into details but only make clear the most important steps. 

Introducing perturbations in the line elements leads to $ds^2=a^2(\eta)\left[-(1+A)d\eta^2+2B_adx^adt+(\tilde\delta_{ab}+E_{ab})dx^adx^b\right]$, with the conformal time defined as $d\eta=dt/a(t)$. Perturbations are encoded in $A,~B_a$ and $E_{ab}$ functions of $(\eta,\vec{x})$, and which are decomposed using the Scalar-Vector-Tensor decomposition (see e.g. Ref. \refcite{peter}). The important point is that the perturbed FLRW space-time is not the FLRW space-time. The two spaces can however be compared so as to identify the perturbative degrees of freedom. Doing this comparison supposes that one can unambiguously establish a correspondance between the points in the FLRW space, and the points in the perturbed FLRW space. However, because of diffeomorphism invariance, one can perform arbitrary changes of the coordinates system of the FLRW space, which may mimick the presence of perturbations. In order to identify the true perturbative degrees of freedom, one searches for gauge-invariant quantities, that is combinations of the Scalar-Vector-Tensor decomposition of $A,~B_a$ and $E_{ab}$ which are invariant by doing a gauge transformation $x^\mu\to x^\mu+\epsilon^\mu(x^\nu)$.

In an Hamiltonian framework with Ashtekar variables, the phase space is extended to incorporate inhomogeneous degrees of freedom: $A^i_a=\gamma\bar{k}\delta^i_a+\delta A^i_a$ (note that $\delta A^i_a=\delta K^i_a+\gamma\delta \Gamma^i_a$), $E_i^a = \bar{p} \delta_i^a +\delta E_i^a$, $N=\bar{N}+\bar{N}\phi$, and $N^a=\bar{N}^a+\partial^aB$. The inhomogeneous degrees of freedom are modelled here as first order perturbations. Background variables describing the homogeneous and isotropic sector are denoted by barred quantities. In the following, we will set $\bar{N}=a$ which select the conformal time as the time variable. From this, vector modes and tensor modes are respectively defined as being transverse, and transverse and traceless. As an example, vector modes contraint the perturbations to satisfy $\delta^i_a\delta E^a_i=0$. A correspondence between perturbations in the Ashtekar formalism and in the standard Lagrangian approach is given in Ref. \refcite{cailleteau_2012}. Similarly, the conditions for selected scalar, vector or tensor modes are given in Eqs. (18), (19) and (20).  Obviously, one should introduce perturbations of the matter content too, $\varphi=\bar\varphi+\delta\varphi$ and $\pi_\varphi=\bar{\pi}_\varphi+\delta\pi_\varphi$. Introducing this perturbed variables in the Hamiltonian constraints and the spatial diffeomorphism constraints lead to\footnote{Note that as usually done in the perturbative expansion, both constraints are expanded up to {\it second} order. However, the second order is only built from the square of {\it first} order perturbations. This is so because the least action principle tells us that the first order is zero on solutions of the zeroth order. one thus needs to go up to second order to get the dynamics of the perturbative degrees of freedom. In the following, we will use the term "second order" in this precise sense (i.e. square of the first order perturbative degrees of freedom and no second order perturbative degrees of freedom).}
\begin{eqnarray}
	S[N]&=&\displaystyle\int_{\Sigma_t}d^3x\left[\bar{N}\left(H^{(0)}+H^{(2)}\right)+\delta N H^{(1)}\right], \\
	D[N^a]&=&\displaystyle\int_{\Sigma_t}d^3x\left[\bar{N}^a\left(D^{(0)}_a+D^{(2)}_a\right)+\delta N^a D^{(1)}_a\right].
\end{eqnarray}
The expression of each term can be found in Refs. \refcite{cailleteau_2012a,cailleteau_2012} with contributions from the gravitational and the matter sectors. The zeroth order simply corresponds to the background dynamics as described previously. They only depend on the background variables and lead to the Friedmann and Klein-Gordon equations for a strictly homogeneous and isotropic universe. Both background and perturbative dynamical variables are involved in the first and second order. However, background variables are not dynamical here, but mere parameters fixed by solving for the zeroth order, and simply encoding the fact that perturbations evolve in a dynamical background space. The first order constraints are related to gauge transformations. Under gauge transformations $x^\mu\to x^\mu+\epsilon^\mu(x^\nu)$, a quantity $X$ is transformed along this vector by $\delta_{\epsilon^\mu}X=\left\{X,S^{(1)}(\bar{N}\epsilon^0)+D^{(1)}(\epsilon^a)\right\}$. In other words, the first order constraints generate the gauge transformations. Moreover, the first order constraints are gauge invariant which translates into the fact that their Poisson brackets are zero\cite{cailleteau_2012a,langlois_1994}. Gauge-invariant variables are finally generated by searching for a canonical transformation from $(\delta A^i_a,\delta E_i^a)$ to $(Q^i_a,P^a_i)$ such as $\delta_{\epsilon^\mu}Q^i_a=\left\{Q^i_a,S^{(1)}(\bar{N}\epsilon^0)+D^{(1)}(\epsilon^a)\right\}=0$. Once the gauge invariant variables are identified, their dynamics are given by $\dot{Q}^i_a=\left\{{Q}^i_a,S(N)+D(N^a)\right\}$. Because the zeroth orders in $S$ and $D$ are functions of the background variables only, their Poisson brackets with the perturbative degrees of freedom is zero. Because the new variables are gauge invariant, their Poisson brackets with the first orders in $S$ and $D$ are zero. The dynamics of the gauge-invariant, perturbative degrees of freedom is thus totally fixed by their Poisson brackets with the second orders in $S$ and $D$.

There is an important point to mention here. If perturbative degrees of freedom are to be described by gauge-invariant variables, it is mandatory to have an Hamiltonian constraint and a spatial diffeomorphism constraint that form a system of first-class constraints. This is because gauge-invariant quantities are defined from gauge transformations generated by $S$ and $D$. For those gauge-invariant variables to remain gauge-invariant accross evolution, it is thus necessary that the generators of the gauge transformations are preserved accross evolution. But in general relativity, this is precisely the generators of gauge transformations which also generates the dynamics. That $S$ and $D$ are preserved accross evolution is guaranteed by the fact they form a system of first-class constraints. That $S$ and $D$ are the generators of gauge transformations in addition, ensures that gauge-invariant variables are preserved accross evolution. This explains why in the classical, general relativistic treatment of cosmological perturbations, the first orders of the constraints, $H^{(1)}$ and $D^{(1)}$, also appear to be gauge-invariant (that is preserved accross time evolution). One also notes that in general relativity, an explicit computation of the Poisson brackets of the constraints expanded up to second order about FLRW space shows that those truncated constraints still form a system of first-class constraints.

\subsection{The deformed algebra approach from an effective perspective}
\label{ssec:da}
Let us return back to the case of the strictly FLRW space. Its loop quantization is performed by expressing the curvature of the Ashtekar connection using holonomies\footnote{The holonomy/flux algebra is quantized and not the connection/densitized triad algebra in LQG.}, $F^k_{ab}\propto\lim_{A\to0}\left(\mathrm{Tr}\left[(h^{(\lambda)}-1)/\lambda^2\right]\right)\propto\bar{k}^2$ with $h^{(\lambda)}$ an holonomy evaluated around a square plaquette with an area $A=\lambda^2$. This curvature is the same at any point of the FLRW space as a result of the homogeneity and isotropy, which is obtained by paving with identical plaquettes. In LQG, the area operator has a discrete spectrum with a minimal area gap given by $\Delta=2\sqrt{3}\pi\ell^2_\mathrm{Pl}\gamma$. Taking this into account means that the curvature should be computed taking $A\to\Delta/\bar{p}^\alpha$, which gives $F^k_{ab}\propto\sin^2(\lambda\bar{k})/\lambda^2$. We note here that $\lambda=\sqrt{A}$ is a function of the phase-space variable, $\bar{p}$, with an ambiguity parameter, $\alpha\in[0,1]$. Chosing $\alpha=1$ here corresponds to the so-called $\bar\mu$-scheme (see Ref. \refcite{bojo_cons} for a discussion). The Hamiltonian constraint is thus expressed using this non-local curvature and subsequently quantized using loops techniques\cite{ashtekar_2003}. Among the different quantum states, one can select states that are sharply peaked around the classical, FLRW solutions in the large volume limit. Their dynamics shows that in the Planck era, the universe goes through a bounce. By finally taking the quantum expectation value of the total Hamiltonian about these states, one ends with an effective, quantum-corrected Hamiltonian: $H_{QC}=-3\sqrt{\bar{p}}/(8\pi G)(\sin\gamma\bar{\mu}\bar{k}/\bar{\mu}\gamma)^2+[\pi_\varphi^2/(2\bar{p}^{3/2})+\bar{p}^{3/2}V(\varphi)]$ (with $\bar\mu=\sqrt{\Delta}/a$). The modified Friedmann equation is deduced from this Hamiltonian, i.e.
\begin{equation}
	\left(\frac{\dot{a}}{a}\right)^2=\frac{8\pi G}{3}\rho\left(1-\frac{\rho}{\rho_c}\right). \label{eq:modfried}
\end{equation}
The dynamics of the bouncing background is discussed in details in another paper of this issue. We just mention here that the bounce occurs for $\rho=\rho_c$ with $\rho_c$ a critical energy density fixed by the minimal area gap, $\Delta$, and the Barbero-Immirzi parameter. The key-point here is that the {\it quantum-corrected} Hamiltonian, $H_{QC}$, provides a very accurate, though effective description of the quantum universe in LQC \cite{ashtekar_2006,rovelli_2014}. The classical limit is recovered independently of the value of $\gamma$. This is easily seen since the classical (or large volume) limit simply means that the area gap, $\Delta$, is much more smaller than $a^2$, leading to $\bar\mu\to0$ irrespectively of the value of $\gamma$.

The goal of the deformed algebra approach is to develop a description of the quantum, perturbed FLRW space-time at an effective level. The idea is to find a prescription to get an effective Hamiltonian up to second order in perturbations which takes into account that the background space is now a quantum one. By inspecting the case of the background, one easily figures out that the effective description is obtained by replacing $\bar{k}$ by $\sin(\gamma\bar{\mu}\bar{k})/\bar{\mu}\gamma$ in the classical Hamiltonian, the spatial diffeomorphism constraint being kept equal to zero because of homogeneity. At the background level, the algebra of constraints remains closed. Incorporating perturbative degrees of freedom can be done by starting from the {\it perturbed} classical Hamiltonian and spatial diffeomorphism constraints, and then make use of prescriptions in line with the one introduced for the background, i.e. $\bar{k}\to\sin(\gamma\bar{\mu}\bar{k})/\bar{\mu}\gamma$. In the zeroth order of the constraints, this exact prescription is used as it comes from the full loop quantization of the FLRW space. The first and second perturbative orders of the constraints are composed of background dynamical variables, and perturbative dynamical variables. How the background variables should be effectively modified so as to take into account that perturbations evolve in a quantum background and not a classical one, is obviously less constrained. It was however rapidly realized that the above-derived prescription yields an algebra of constraints up to second order that is not closed anymore (note that the matter content plays a key-role in the presence of anomalies), i.e. $\{ \mathcal{C}^{QC}_I, \mathcal{C}^{QC}_J \} = {f^K}_{IJ}(A^j_b,E^a_i) \mathcal{C}^{QC}_K+\mathcal{A}_{IJ}$ with $\mathcal{A}_{IJ}$ some anomalous terms (the $QC$ superscript means quantum-corrected at an effective level). Those anomalous terms come from Poisson brackets between the zeroth order in the constraints with the second order, or between two first orders in the constraints. Taking lessons from general relativity, one should search for a prescription, $\bar{k}\to g(\bar{k})$, such as the algebra of constraints up to second order remains closed\footnote{Note that the functions $g$ are not totally free. They are chosen to be pseudoperiodic functions of $\bar{k}$, e.g. $\bar{k}\to\sin(n\gamma\bar{\mu}\bar{k})/n\bar{\mu}\gamma$ with $n\in\mathbb{N}$, in the spirit that the effective description should capture the use of holonomies of the connection, and not the connection itself. Some counterterms are also added to cancel the anomalies.}. The price to pay to cancel the anomalous terms is that the structure functions of the algebra, ${f^K}_{IJ}$, may also gain quantum corrections and thus differ from what is obtained in classical general relativity. A detailed studies of this procedure was done for vector and scalar modes \cite{mielc_2012,cailleteau_2012b} without assuming a specific quantization scheme for the background, i.e. $\bar{\mu}=\sqrt{\Delta/\bar{p}^\alpha}$. This requires tedious calculations. Surprisingly enough, requiring that the algebra is closed, i.e. $\mathcal{A}_{IJ}=0$, as well as the recovery of the classical constraints in the large volume limit leads to a {\it unique} solution for the quantum-corrected constraints up to second order. Interestingly too, this approach fixes the quantization scheme to be the $\bar\mu$-scheme, i.e. $\alpha=1$. Finally, this unique solution appears rather simple since the closure of the quantum-corrected algebra, or {\it deformed algebra}, reads\cite{cailleteau_2012}
\begin{eqnarray}
	\left\{D^{QC}[M^a],D^{QC}[N^a]\right\}&=&D^{QC}\left[M^b\partial_bN^a-N^b\partial_bM^a\right], \\
	\left\{D^{QC}[M^a],S^{QC}[N]\right\}&=&S^{QC}\left[M^b\partial_bN-N\partial_bM^b\right], \\
	\left\{S^{QC}[M],S^{QC}[N]\right\}&=&\mathbf{\Omega}D^{QC}\left[q^{ab}(M\partial_bN-N\partial_bM)\right],	
\end{eqnarray}
with $\mathbf{\Omega}=\cos(2\gamma\bar\mu\bar{k})=1-2\rho/\rho_c$, which only depends on the background variables. We stress that for $\rho>\rho_c/2$ (that is close to the bounce), $\mathbf{\Omega}$ becomes negative-valued. (We will comment on that later on.)

Once the quantum-corrected constraints have been defined and built to close the (now deformed) algebra, the procedure follows the path of general relativity. At the zeroth order, one recovers the standard results obtained in the effective treatment of LQC, Eq. (\ref{eq:modfried}). The first order is used to define the gauge-invariant variables. We note that gauge transformations are quantum corrected. This is however not a surprise since in general relativity, dynamics is part of the gauge transformations. If the dynamics receives quantum corrections, one could expect that gauge transformations receive corrections too. Considering a universe filled with a single scalar fields, there is one gauge-invariant scalar degree of freedom, and two gauge-invariant tensor degrees of freedom. They are defined in terms of Mukhanov-Sasaki variables, $v_\mathrm{S}=z_\mathrm{S}\mathcal{R}$ with $\mathcal{R}$ the 3-dimensional scalar curvature, and $v_{\mathrm{T},s}=z_\mathrm{T}h_s$ with $h_s$ ($s=1,~2$) the helicity states of gravitational waves. The functions $z_i$ are expressed using background variables: $z_\mathrm{S}=a^2\bar\varphi'/a'$, and $z_\mathrm{T}=a/\sqrt{\mathbf{\Omega}}$. Their respective equation of motion in spatial Fourier space takes the form of a Schr\"odinger equation with a time-dependent frequency
\begin{equation}
	v''_i+\left(\mathbf{\Omega} k^2-\frac{z''_i}{z_i}\right)v_i=0,
\end{equation}
with $k=|\vec{k}|$ the wavenumber of the spatial eigenmodes, and $i$ standing for $\mathrm{S}$ or $\mathrm{T},s$. The time-dependant potential-like term, ${z''_i}/{z_i}$, encodes the impact of the dynamical background on the propagation of perturbations.

\subsection{Comments}
Let us briefly comment on the above construction. We refer to Refs. \refcite{barrau_ijmpd,barrau_jcap} for detailed discussions.

First, we remind that the above approach is a way to implement at an effective level the impact of a quantum background on the propagation of inhomogeneities treated as perturbations. Special attention is paid to the closure of the algebra of constraints so as to ensure that the perturbative degrees of freedom can be described by gauge-invariant variables, in the spirit of Dirac viewpoints that observables (and certainly cosmic inhomogeneities are observables) should be gauge-invariant. However the very quantum nature of the background is only treated at an effective level, but consistently so with respect to the issue of gauge-invariance. This is also motivated by the fact that in general relativity this is the same constraints that generate both gauge transformations and the dynamics. In that sense, if the dynamics is quantum-corrected, then the gauge transformations are corrected too, and both should be treated on an equal footing. \\

Second, it is worth to mention that the same results as described in Sec. \ref{ssec:da} have been obtained using different paths. In particular, this has been obtained as the effective description of a fully quantized model built from homogeneous and isotropic cells, inhomogeneities being modelled as the weak interaction between the different cells\cite{wilson_2012a}. Similarly, the same description of perturbations is obtained by directly perturbing the effective, quantum-corrected FLRW Hamiltonian\cite{wilson_2012b} (as opposed to the above-described approach, which implements effective corrections to the classical, perturbed Hamiltonian). \\

Third, one should not be surprised that the deformation of the algebra appears only once perturbations are switched on. Note first that the deformation is in the last Poisson bracket and appears as a multiplicative function to the spatial diffeomorphism constraint in the right-hand-side. At the background level however, this constraint is zero because of homogeneity. This trivially solves for the last Poisson bracket whatever a possible multiplicative deformation would be. \\

Finally, the fact that $\mathbf{\Omega}$ becomes negative close to the bounce when $\rho>\rho_c/2$ has impressive consequences. In the evolution of perturbations, the deformed algebra is made manifest in the $\mathbf{\Omega}k^2$ terms which leads to a sort of tachyonic instability. The equation of motion for the perturbations is {\it hyperbolic} during the classical contraction and the classical expansion, while it becomes {\it elliptic} around the bounce. 

But the most impressive consequence is obtained by a direct inspection of the deformed algebra of constraints: the transition from a positive-valued $\mathbf{\Omega}$ to a negative-valued one can be interpreted as a change of signature\cite{paily}. (Further arguments are given in Ref. \refcite{barrau_jcap} to show that the change from an hyperbolic equation for the perturbations to an elliptic one is the mark of a true change of signature.) This is surprising at first sight. For example, Riemannian, 4-dimensional spaces (for which the undeformed algebra of hypersurface deformation is associated to) are known to have a fixed signature. However, this simply means that the potential 4-dimensional structure describing the {\it effective} space associated to the deformed algebra of hypersurface deformation is not a Riemannian space (though tentative geometrical interpretations of the 4-dimensional, effective space are still unknown yet). Stated otherwise, since covariance of the solutions of Einstein's equation is represented in an Hamiltonian framework by the undeformed algebra of hypersurface deformation, this resulting deformed algebra does not represent changes of coordinate system anymore. This obviously makes the physical interpretation of the effective space difficult, and considering evolution in a universe which experienced some Euclidean phase is clearly odd. However, this in a sense is a limitation if one is willing to make use of 4-dimensional notions as e.g. invariant line element. Nonetheless, the deformed algebra still defines clear notions of {\it canonical} observables that are gauge-invariant, and approaches to define notions such as light cone from characteristics of wave equations have been proposed\cite{bojo_jakub}.

\section{Primordial power spectra}
\label{sec:pk}
With a theory of cosmological perturbations at hand, it is possible to make predictions on astronomical observables. We will focus here on a scenario where the universe is filled with a massive scalar field. In this setting, most of the trajectories for the LQC universe are such that the bounce is followed by a phase of classical inflation\cite{linsefors_2013,sloan_2010}. As already explained in our introduction, the observables to be predicted are the statistical properties of the scalar and tensor inhomogeneities at the end of the inflationary era. These can subsequently be translated in observables such as the CMB anisotropies. We will focus here on the primordial power spectra though higher order statistics could be considered.

To this end, one should first define some initial conditions for both the background and the perturbations. In LQC, the bounce could appear as a natural choice since it corresponds to the deep quantum regime. In the deformed algebra approach however, it is not possible to properly define a quantum states for the perturbations at the bounce. Alternatively, if causal evolution is to be taken seriously, the bounce is preceded by a contracting phase and initial conditions are set in the remote past during the classical contraction. Here we focus on the latter option to highlight how the statistical properties of the primordial inhomogeneities are derived. In the remote past, the equation of motion for perturbations is hyperbolic. Similarly, this equation is hyperbolic at the end of inflation after the bounce. Initial conditions and final outputs are thus computed in the Lorentzian regions of the effective space where evolution is clearly defined. However, perturbations will experienced a phase of instability around the bounce which can be viewed as the analog of a supersonic flow, while initial and final conditions would be set in the subsonic part of the flow. In this setting, the conformal time remains an appropriate parameter to label trajectories during the Euclidean phase. It is worth to mention here that apart from the initial state chosen for the perturbations, this is the background evolution which will set up the shape and the amplitude of the $\mathcal{P}_i(k)$'s through $z_i$ and $\mathbf{\Omega}$. There are two kinds of relevant parameters for the background evolution. First the critical energy density $\rho_c$ that is fixed to its standard value $\sim0.4~m^4_\mathrm{Pl}$, and for the case of a massive scalar field, its mass $m$. These are fundamental parameters. Second, the relative amount of potential energy in the scalar field at the bounce, $x_\mathrm{B}$, will set up the number of e-folds during inflation\footnote{Note that this is the only free parameter since at the bounce $\rho=\rho_c$ and the Hubble rate is zero.}. Chosing $x_\mathrm{B}$ means that a single trajectory for the background is picked up, to which correspond a single amount of e-folds during inflation. The parameter $x_\mathrm{B}$ is unequivocally related to the relative phase between $\varphi$ and $\dot{\varphi}$ in the contracting phase, called $\theta$ (see Refs. \refcite{linsefors_2013,bolliet_2015,barrau_ijmpd}). \\

Detailed analytical and numerical computations of the $\mathcal{P}_i(k)$ for the tensor modes and the scalar modes in the deformed algebra approach, and setting the initial conditions in the remote past, can be found in Refs. \refcite{bolliet_2015} \& \refcite{schander}, respectively. Though there are differences between the two types of modes, their treatment is rather similar at a qualitative level. We will thus focus on the case of tensor modes which is most easily understood.

In the remote past, the evolution is classical and the curvature radius of the universe tends to infinity with $\mathbf{\Omega}\to1$. This means that the equation of motion for the Mukhanov-Sasaki variables  reads $v''_\mathrm{T}+k^2v_\mathrm{T}=0$. The standard Minkowski vacuum can be chosen as initial conditions for the perturbations, which is characterized by the mode functions $v_\mathrm{T}(\eta\to-\infty)\simeq\exp(-ik\eta)/\sqrt{2k}$. To understand the final shape of the primordial power spectrum as a function of scales, one should have a clear inspection of the time-dependant frequency, $\omega^2(\eta)=\mathbf{\Omega}k^2-z''_\mathrm{T}/z_\mathrm{T}$, and more specifically how its sign evolves as a function of the conformal time, and as a function of scales. This is the relevant parameter at a qualitative level since a mode at given $k$ is amplified if the square of its time-dependant frequency is negative, while the mode is in a more standard oscillatory regime for positive square of the frequency. Initially, all the modes are in an oscillatory regime. Then the time and duration at which the modes enter in the amplified regime will set up the shape of the spectrum. 

Three different regions for the scale behaviour of $\mathcal{P}_\mathrm{T}(k)$ have been identified, which reflect the three important phases in the background evolution (i.e. the classical contraction, the bounce, and the classical expansion)\cite{bolliet_2015}. One first notes that up to the onset of inflation (which comes shortly after the bounce), the term $z''_\mathrm{T}/z_\mathrm{T}$ is bounded from above with its maximal value attained at the bounce, $\mathrm{max}\left|z''_\mathrm{T}/z_\mathrm{T}\right|=40\pi G\rho_c$. This set an important comoving scale dubbed $k_\mathrm{UV}=\sqrt{40\pi G\rho_c}$. All the modes with $k>k_\mathrm{UV}$ (i.e. the smallest scales) are basically described by $v''_\mathrm{T}+\mathbf{\Omega} k^2v_\mathrm{T}=0$ from the contraction up to the onset of inflation. Since during the Euclidean phase $\mathbf{\Omega}<0$, the amplitude of these modes grows exponentially during this phase, i.e. $v(\eta_i)\sim\exp(k\int_{\Delta\bar\eta}\sqrt{\left|\mathbf{\Omega}(\bar\eta)\right|}d\bar\eta)\times\exp(-ik\eta_i)$, with $\eta_i$ the onset of inflation and $\Delta\bar\eta$ denoting the width of the Euclidean phase\footnote{We use $\bar\eta$ instead of $\eta$ during the Euclidean phase to emphasize that the "conformal time" is used as a parameter labelling trajectories and not as a proper time coordinate.}. Though during inflation these modes will becomes superhorizon, i.e. $k^2<\left|z''_\mathrm{T}/z_\mathrm{T}\right|$ (note that $\mathbf{\Omega}\simeq1$ during inflation), their amplitude is mainly set by the exponential growth during the Euclidean phase. As a consequence, the primordial power spectrum grows exponentially with $k$ in the ultraviolet regime, $k>k_\mathrm{UV}$. It scales as $\mathcal{P}_\mathrm{T}(k>k_\mathrm{UV})\sim\exp(2k/k_\mathbf{\Omega})$ with $k_\mathbf{\Omega}=1/\int_{\Delta\bar\eta}\sqrt{\left|\mathbf{\Omega}(\bar\eta)\right|}d\bar\eta$. It has been shown that $k_\mathbf{\Omega}\propto\sqrt{G\rho_c}$ with a numerically computed prefactor making $k_\mathbf{\Omega}$ of the same order of $k_\mathrm{UV}$. Since a negative $\mathbf{\Omega}$ is the sign of the Euclidean phase (or at least of the instability), the small scales are strongly impacted by the Euclidean phase.

For $k<k_\mathrm{UV}$, this means that $k<k_\mathbf{\Omega}$ and the exponential growth due to the Euclidean phase is of order unity. These modes are then poorly affected by $\mathbf{\Omega}$ and in that sense, they are said to be unaffected by the Euclidean phase. For these larger scales, two regions are identified. The first region is in the infrared regime. This corresponds to scales $k<k_\mathrm{IR}$ with $k_\mathrm{IR}\propto\left(m^2\sqrt{G\rho_c}/\left|\cos\theta\right|\right)^{1/3}$, which is strictly smaller than $k_\mathrm{UV}$. The IR modes enter the amplification regime during the classical contraction. The power spectrum has been shown to be nearly scale invariant (with a slightly red tilt) for these IR scales. It is analytically shown that it roughly scales as $\mathcal{P}_\mathrm{T}(k\ll k_\mathrm{IR})\sim(m/\cos\theta)^2\times\ln^2(m/\sqrt{G\rho_c})$\cite{bolliet_2015}.

Finally, modes with scales defined by $k_\mathrm{IR}<k<k_\mathrm{UV}$ does not feel the curvature of the universe during the contraction. As a consequence, they enter in the amplification regime during the bounce, and then during inflation. Analytical calculations are made difficult in this regime. Numerical results then show that the power spectrum presents damped oscillations. The maximal amplitude of these oscillations are about one to two orders of magnitude of the average. This damped, oscillatory regime extends on about two to three decades in scales.

\begin{figure}
\begin{center}
	\includegraphics[scale=0.425]{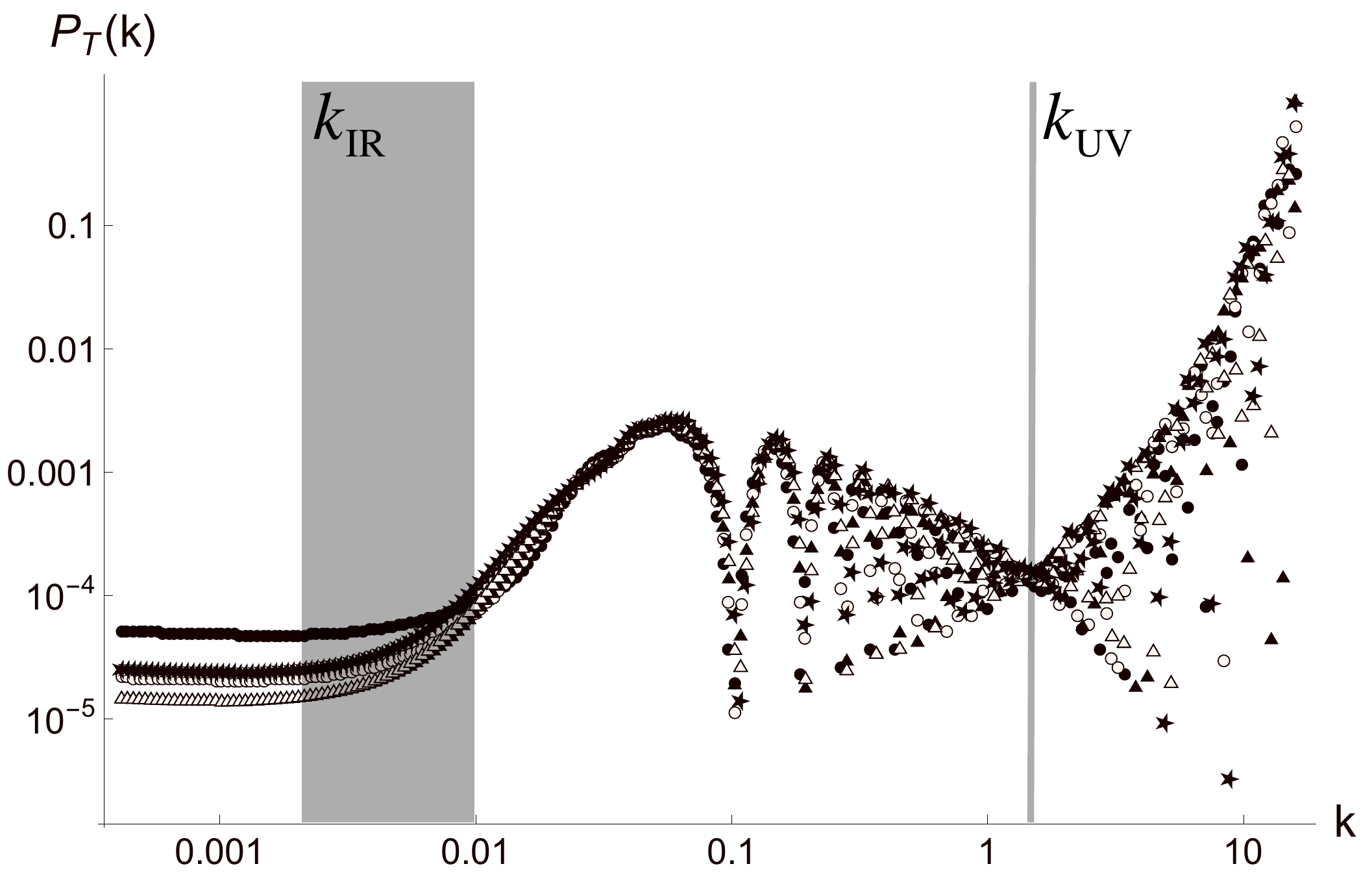}
	\caption{Primordial power spectrum for the tensor modes as predicted in the deformed algebra approach, setting the initial conditions in the contracting phase\cite{bolliet_2015}. The different curves correspond to different values of $\theta$ (corresponding to different relative amount of potential energy in the scalar field at the bounce). This mainly affects the infrared transition scales, $k_\mathrm{IR}$, and the amplitude of the spectrum for $k<k_\mathrm{IR}$.}
	\label{fig:pk}
\end{center}
\end{figure}

An example of the numerical results is depicted on Fig. \ref{fig:pk} for the tensor modes, and for different values of $\theta$. The three regions are easily recognized (note that this is comoving wavenumbers). We stress that the qualitative shape of these primordial power spectra exhibiting three regions is rather independent of the parameters $\rho_c,~m$ and $\theta$. The precise values set however the amplitude of the spectra in the different regions as well as the value of the transition scales, $k_\mathrm{IR}$ and $k_\mathrm{UV}$.

For the scalar modes, it was numerically shown that the damped oscillatory regime at intermediate scales, and the exponential growth in the UV are preserved\cite{schander}. In the IR (large scales) however, the numerical results show a blue spectrum scaling as $\propto k^3$. (We refer to Ref. \refcite{schander} for details.) \\

Let us briefly comments on those results. First, the exponential growth in the UV is the mark of the Euclidean phase that translates in an instability in the equation of motion of the perturbations. In the IR and oscillatory regimes, the primordial power spectra are not affected by this instability. This is strengthened by the fact that the primordial power spectra predicted {\it without} the $\mathbf{\Omega}$-instability\footnote{Here in the simple meaning that $\mathbf{\Omega}$ is set equal to one throughout the entire, bouncing cosmic history.} is basically identical to the ones showed here for $k<k_\mathrm{UV}$ \cite{bolliet_2015}. In the UV regime and in absence of the $\mathbf{\Omega}$-instability, modes are amplified during inflation only and one recovers the standard nearly scale invariant power spectrum as predicted by inflation, with the proper slighty red tilt. (This is relevant for the so-called dressed-metric approach for modelling cosmological perturbations in LQC \cite{aan1,aan2,aan3}. Initial conditions are set in the contracting phase in Ref. \refcite{bolliet_2015} while they are set at the bounce in Ref. \refcite{aan3}.)

This nonetheless raises some questions. Even if one is convinced that there is indeed a change of signature, the UV exponential growth may still points towards the non applicability of the linear perturbative treatment at those small scales. This however does not mean that the deformed algebra is inherently faulty. First, the quantum modelling of the perturbed FLRW space-time through patches\cite{wilson_2012a} suggests that the equation of motion of perturbations as derived in the deformed algebra approach should simply not be used down to arbitrary small scales, though it remains valid for larger scales. Second, the change of signature leads to an exponential growth of the power in the UV because initial conditions are set in the contracting phase, for this leads to solve for an {\it initial value} problem with an equation of motion which becomes elliptic. Alternatively, one could set initial conditions at the {\it silent} hypersurface ($\mathbf{\Omega}=0$) of the expanding branch\cite{jakub_silent}. Also, one could instead solve for a {\it boundary value} problem in the deep quantum regime around the bounce where the equation of motion is elliptic. That allows for avoiding instabilities though it raises some conceptual issues regarding causality which is abandoned here\cite{bojo_jakub}. 

\section{Consequences on the cosmic microwave background}
\label{sec:cl}
The model described above and used to derive the primordial power spectra is obviously not free of assumptions. At a fundamental level, the deformed algebra approach does not consider the full quantum features of the quantum FLRW background. It however consistently addresses the gauge issue for perturbations at an effective level. For the matter content then, one simply considers a massive scalar field. For the specific calculations of the primordial power spectra, one assumes a truly causal evolution and thus the initial conditions are set in the classical contracting branch. In addition, no backreaction are taken into account here. Each of those assumptions could be (and are) criticized. However, each of them also appears as either quite natural, or quite simple. The key-point here is that despite all the arguments in favor of this, and despite all the criticisms against this specific model, one has predictions at hand and this can be compared to the observational data. This has been done in Ref. \refcite{bolliet_2016} and we summarize here their argument. \\

The CMB anisotropies allows for probing scales from $k\sim10^{-4}$ Mpc$^{-1}$ to $k\sim1$ Mpc$^{-1}$. Thanks to the high precision measurements of the Planck satellite in temperature, the primordial power spectrum of scalar modes is now well constrained in these scales. The $B$-modes of polarized anisotropies is only upper bounded. This (with temperature measurements) leads to an upper bound on the tensor modes roughly in the first two decades of the aforementioned $k$ range with $\mathcal{P}_\mathrm{T}\lesssim10^{-10}$. The primordial power spectra shown before as a function of {\it comoving} wavenumbers can only be probed in a restricted range of $k$. The remaining question is thus to know on which range of comoving wavenumbers the physical range of observed wavenumbers falls. This obviously depends on the total amount of e-folds during inflation. Assuming the standard value of $\rho_c\simeq0.4~m^4_\mathrm{Pl}$, this is the parameter $\theta$ (or equivalently the relative amount of potential energy at the bounce) that mainly drives the total number of e-folds (the mass of the scalar field having only a tiny impact). Once the parameter $\theta$ is set to a given value, this first uniquely defines the amplitude of the power spectrum as well as $k_\mathrm{IR}$ and $k_\mathrm{UV}$. Second, this uniquely defines the total number of e-folds and thus where the observational window of scales falls in the primordial power spectrum\cite{bolliet_2016}. Roughly speaking, everything is fixed once a value of $\theta$ has been arbitrarily chosen. This parameter $\theta$ cannot be known a priori and one should explore its entire range, $\theta\in[0,2\pi]$. 

There are roughly three configurations. First, it was shown that most of the range of values of $\theta$ leads to a total number of e-folds during inflation much greater than 60\cite{linsefors_2013}. This means that scales are so redshifted that the observable window falls in the deep UV region of the spectrum. This corresponds to the exponential growth of the spectrum, which leads to a predicted $\mathcal{P}_\mathrm{T}$ much greater than the upper bound of $10^{-10}$.

Second, one has to explore values of $\theta$ that are fine-tuned to get a much smaller amount of e-folds during inflation. In that spirit, it could be so that $\theta$ is such that the observable window falls in the IR part of the spectrum. However to do so, the precise dynamics of the bouncing universe is such that it goes through a phase of exponential contraction --called {\it deflation}-- prior to the bounce\cite{linsefors_2013}. This boosts up the amplitude of the primordial power spectrum in the IR way above the upper bound of $10^{-10}$. 

Finally, the parameter $\theta$ could be so fine-tuned that the observable window falls exactly on the oscillatory regime. In this case too however, the amplitude of the oscillation is such that the primordial power spectrum is above its upper bound. \\

To summarize, whatever the evolution of the universe is, the predicted angular power spectrum for the $B$-modes of polarized CMB anisotropies is above the current upper bound, and this specific model is excluded by observations. (The case of scalar modes which mainly drives the angular power spectra in temperature and $E$-modes would strengthened this conclusion.)

Does this mean that LQG is excluded? Certainly not since LQC is based on minisuperspace quantization inspired by LQG, but it is not yet {\it deduced} from LQG. Does this mean that LQC is excluded? Certainly not since the primordial power spectra (as astronomical observables) are predicted in a specific theoretical framework for perturbations in a quantum background, namely the deformed algebra approach. There exists alternative ways for constructing the theory of cosmological perturbations in LQC, and their predicted primordial power spectra differ from the ones derived here, especially in the UV regime\cite{aan3,bolliet_2015}. Does this mean that the deformed algebra is excluded? The answer is again in the negative. As already explained, the shape of the primordial power spectra presented above (and which are obviously excluded by observations) does not solely rely on the deformed algebra framework, but also on the specific choice to set the initial conditions in the contracting phase {\it and} to solve for an initial value problem (and not a boundary value problem).

Despite all these precautions, this shows that a consistent scenario for modelling the cosmic history in LQC is excluded by observations. Though restricted to a specific version of LQC, this concrete examples highlight the predictive power of the approach.


\section{Conclusion}
\label{sec:concl}
Since the seminal papers on LQC in the late nineties, the field has dramatically been developed. In particular, it was rapidly realized that adding inhomogeneities would be a key advance for the field since this is mandatory to make predictions to be subsequently compared to astronomical observations. Despite a consensus concerning the background, there are still some debates on how inhomogeneities should be modelled. The deformed algebra approach described here consistently implements the quantum corrections from the viewpoint of gauge issues. However, these quantum corrections are considered at an effective level. This approach is obviously not free of any assumptions, but in this precise setting for treating cosmic inhomogeneities in the quantum era (with initial conditions put in the contracting phase), predictions on the angular power spectra of the CMB anisotropies can be purchased, and as recently shown, the predicted $B$-modes is above the current, observational upper bound. This emphasizes that the approach is falsifiable! 

That this specific version of LQC is excluded does not exclude however LQG nor LQC, but only the {\it conjunction} of the deformed algebra approach {\it and} initial conditions of the perturbations in the contracting phase. It is thus important to remind clearly what the assumptions are. First the universe is filled with a massive scalar field and described at an effective level with holonomy corrections only. Second, cosmic inhomogeneities are modelled as perturbations --with no backreaction-- in the deformed algebra approach. Third, a true causal evolution is assumed. This means that initial conditions are set in the contracting phase and one solves for an initial value problem. Four, the initial state for perturbations is assumed to be the Minkowski vacuum. \\

Alternative models for cosmic inhomogeneities exist, and some of them have been pushed to predict observables. As an example, the dressed metric approach is a minisuperspace viewpoint where the perturbed FLRW metric is quantized. The phase-space is the product of the standard phase-space for background variables times the phase-space of the {\it classical}, gauge-invariant variables for perturbations. Cosmic inhomogeneities here are also modelled as perturbations. In this setting, primordial power spectra\cite{aan3,bolliet_2015,agullo_2015} have been derived. Predictions about higher order statistics (i.e. bispectrum) also exists\cite{agullo_2015b}. \\

Let us finally mention a very last point. The current most advanced models of cosmic inhomogeneities in LQC are the deformed algebra and the dressed metric approaches. Both of them assume that the underlying starting point is a perturbed FLRW space. From a cosmological viewpoint however, the cosmological principle can be stated as a statistically homogeneous and isotropic universe. A perturbative FLRW space is sufficient to realize this, but not necessary. Obviously the mathematical coherence of considering cosmic inhomogeneities as perturbative degrees of freedom can be checked, or at least controlled through regularizations even in the LQC context\cite{aan2}. Nonetheless, this is a restriction of the models and one may simply miss that close to the bounce, inhomogeneities are not described by perturbative degrees of freedom (though it is worth to mention that the perturbative viewpoint has an advantage regarding the very tiny amount of non-gaussianities as observed in the CMB). This is very reminiscent to the issue of cosmic inhomogeneities in the late time universe that may not be considered as perturbations anymore because of the very high density contrast, though in LQC this would be due to the high density. 

In that respect, the so-called {\it hybrid models} developed in e.g. Ref. \refcite{gomar_2015} takes into account backreaction effects, though pratically speaking, the equations are solved in the limit of a negligible backreaction. 

Still in the framework of minisuperspace quantization, it could be useful to develop some LQC-like quantization of non-perturbative models of the statistically homogeneous and isotropic space-time\cite{inhomo} (such as Swiss-Cheese models\cite{fleury}).

Finally, this question of how to mathematically implement the idea of a statistically homogeneous and isotropic space-time (that is the question of inhomogeneities, either in standard cosmology or in LQC), invites to properly phrase the cosmological principle in the language of LQG. Attempts have been made in this way through spinfoam cosmology\cite{vidotto1,vidotto2}, symmetry-reduced loop quantum gravity\cite{alesci}, and group field theory\cite{gielen}.



\begin{thebibliography}{0}    

\bibitem{barrau_2014} A. Barrau, T. Cailleteau, J. Grain and J. Mielczarek, {\it Class. Quantum Grav.} {\bf 31} (2014) 053001
\bibitem{hawking_1973} S. Hawking and G. F. R. Ellis, {\it The Large Scale Structure of Space-Time} (Cambridge University Press, Cambridge, 1973)
\bibitem{bojowald_2008} M. Bojowald, {\it Living Reviews in Relativity} {\bf11} (2008) 10.12942
\bibitem{ashtekar_2003} A. Ashtekar, M. Bojowald and J. Lewandowski, {\it Adv. Theor. Math. Phys.} {\bf7} (2003) 233
\bibitem{linsefors_2013} L. Linsefors and A. Barrau, {\it Phys. Rev. D} {\bf 12} (2013) 123509
\bibitem{bolliet_2015} B. Bolliet, J. Grain, C. Stahl, L. Linsefors and A. Barrau, {\it Phys. Rev. D} {\bf91} (2015) 084035
\bibitem{sloan_2010} A. Ashtekar and D. Sloan, {\it Phys. Lett. B} {\bf694} (2010) 108
\bibitem{planck_gauss} Planck collaboration, arXiv:1502.01592 [astro-ph.CO]
\bibitem{zaldarriaga_1997} U. Seljak and M. Zaldarriaga, {\it Phys. Rev. Lett.} {\bf78} (1997) 2054
\bibitem{planck_inf} Planck collaboration, arXiv:1502.02114 [astro-ph.CO]
\bibitem{planckbicep} Bicep2/Keck and Planck collaborations, {\it Phys. Rev. Lett.} {\bf114} (2015) 101301
\bibitem{barrau_ijmpd} A. Barrau and B. Bolliet, {\it IJMPD} same issue, arXiv:1602.04452 [gr-qc]
\bibitem{barrau_jcap} A. Barrau, M. Bojowald, G. Calcagni, J. Grain and M. Khagan, {\it JCAP} {\bf05} (2015) 051
\bibitem{iv_bojo} M. Bojowald, G. M. Hossain, M. Kagan and S. Shankaranarayanan, {\it Phys. Rev. D} {\bf79} (2009) 043505
\bibitem{iv_tom} T. Cailleteau, L. Linsefors and A. Barrau, {\it Class. Quant. Grav.} {\bf31} (2014) 125011
\bibitem{adm} R. Arnowitt, S. Deser and C. Misner, {\it Phys. Rev.} {\bf116} (1952) 1322
\bibitem{ashtekar_1986} A. Ashtekar, {\it Phys. Rev. Lett.} {\bf57} (1986) 2244
\bibitem{thiemann} T. Thiemann, {\it Modern Canonical Quantum General Relativity} (Cambridge University Press, Cambridge, 2007)
\bibitem{hojman_1976} S. A. Hojman, K. Kucha\v{r} and C. Teitelboim, {\it Ann. Phys.} {\bf96} (1976) 88
\bibitem{langlois_1994} D. Langlois, {\it Class. Quantum Grav.} {\bf 11} (1994) 389
\bibitem{cailleteau_2012a} T. Cailleteau and A. Barrau, {\it Phys. Rev. D} 85 (2012) 123534
\bibitem{cailleteau_2012} T. Cailleteau, A. Barrau, J. Grain and F. Vidotto, {\it Phys. Rev. D} {\bf86} (2012) 087301
\bibitem{peter} P. Peter and J.-P. Uzan, {\it Cosmologie Primordiale} (Edition Belin, Paris, 2005)
\bibitem{bojo_cons} M. Bojowald, {\it Class. Quant. Grav.} {\bf26} (2009) 075020
\bibitem{ashtekar_2006} A. Ashtekar, T. Pawlowski and P. Singh, {\it Phys. Rev. D} {\bf73} (2006) 124038
\bibitem{rovelli_2014} C. Rovelli and E. Wilson-Ewing, {\it Phys. Rev. D} {\bf} {\bf90} (2014) 023538
\bibitem{mielc_2012} J. Mielczarek, T. Cailleteau, A. Barrau and J. Grain, {\it Class. Quantum Grav.} {\bf 29} (2012) 085009
\bibitem{cailleteau_2012b} T. Cailleteau, J. Mielczarek, A. Barrau and J. Grain, {\it Class. Quantum Grav.} {\bf 29} (2012) 095010
\bibitem{wilson_2012a} E. Wilson-Ewing, {\it Class. Quantum Grav.} {\bf 29} (2012) 215013
\bibitem{wilson_2012b} E. Wilson-Ewing, {\it Class. Quantum Grav.} {\bf 29} (2012) 085005
\bibitem{paily} M. Bojowald and G. M. Paily, {\it Phys. Rev. D} {\bf86} (2012) 104018
\bibitem{bojo_jakub} M. Bojowald and J. Mielczarek, {\it JCAP} {\bf08} (2015) 052
\bibitem{schander} S. Schander, A. Barrau, B. Bolliet, J. Grain, L. Linsefors and J. Mielczarek, {\it Phys. Rev. D} {\bf93} (2016) 023531
\bibitem{aan1} A. Ashtekar, I. Agullo and W. Nelson, {\it Phys Rev. Lett.} {\bf109} (2012) 251301
\bibitem{aan2} A. Ashtekar, I. Agullo and W. Nelson, {\it Phys Rev. D} {\bf87} (2013) 043507 
\bibitem{aan3} A. Ashtekar, I. Agullo and W. Nelson, {\it Class. Quantum Grav.} {\bf 30} (2013) 085014
\bibitem{jakub_silent} J. Mielczarek, A. Barrau and L. Linsefors, arXiv:1411.0272 [gr-qc]
\bibitem{bolliet_2016} B. Bolliet, A. Barrau, J. Grain and S. Schander, submitted to {\it Phys. Rev. D}, arXiv:1510.08766 [gr-qc]
\bibitem{agullo_2015} I. Agullo and N. A. Morris,  arXiv:1509.05693 [gr-qc]
\bibitem{agullo_2015b} I. Agullo, {\it Phys. Rev. D} {\bf92} (2015) 064038
\bibitem{gomar_2015} L. C. Gomar, M. Martin-Benito and G. Mena Marugan, {\it JCAP} {\bf06} (2015) 045
\bibitem{inhomo} A. Krasinski, {\it Inhomogeneous Cosmological Models} (Cambridge University Press, Cambridge, 1997)
\bibitem{fleury} P. Fleury, {\it Light Propagation in Inhomogeneous and Anisotropic Cosmologies}, Ph.D. Thesis (Universit\'e Pierre et Marie Curie, Paris, 2015) arxiv:1511.03702 [gr-qc]
\bibitem{vidotto1} C. Rovelli and F. Vidotto, {\it Class. Quant. Grav.} {\bf25} (2008) 225024
\bibitem{vidotto2} E. Bianchi, C. Rovelli and F. Vidotto, {\it Phys. Rev. D} {\bf82} (2010) 084035
\bibitem{alesci} E. Alesci and F. Cianfrani, {\it Phys. Rev. D} {\bf92} (2015) 084065
\bibitem{gielen} S. Gielen, D. Oriti and L. Sindoni, {\it Phys. Rev. Lett.} {\bf111} (2013) 031301
\end{thebibliography}
\end{document}